\documentclass[fleqn,usenatbib]{mnras}
\usepackage{newtxtext,newtxmath}

\usepackage{ae,aecompl}
\usepackage{graphicx}	
\usepackage{url}
\usepackage{amsfonts}
\usepackage{float}
\usepackage{color}
\usepackage{ulem}

\title[M/L gradients and M/L-colour relations]
{SDSS-IV MaNGA: Stellar M/L gradients and the M/L-colour relation in galaxies}
\author[J. Q. Ge et al.]{
Junqiang Ge, $^1$\thanks{E-mail: jqge@nao.cas.cn}
Shude Mao, $^{2,1}$
Youjun Lu, $^{1,3}$
Michele Cappellari, $^{4}$
Richard J. Long, $^{2,5}$ 
\newauthor
Renbin Yan $^{6}$ \\
$^{1}$National Astronomical Observatories, Chinese Academy
  of Sciences, 20A Datun Road, Beijing 100101, China \\
$^{2}$Department of Astronomy,
               Tsinghua University, Beijing, 100084, China \\
$^{3}$School of Astronomy and Space Science, 
               University of Chinese Academy of Sciences, Beijing 100049, China \\
$^{4}$Sub-Department of Astrophysics, Department of Physics, 
               University of Oxford, Denys Wilkinson Building, Keble Road, 
              Oxford, OX1 3RH, UK \\	  
$^{5}$Jodrell Bank Centre for Astrophysics,
               Department of Physics and Astronomy, The University of Manchester, Oxford Road,
               Manchester M13 9PL, UK \\     
$^{6}$Department of Physics and Astronomy, University of Kentucky, 
               505 Rose Street, Lexington, KY 40506, USA \\
}

\date{Accepted XXX. Received YYY; in original form ZZZ}

\pubyear{2020}

\begin{document}
\label{firstpage}
\pagerange{\pageref{firstpage}--\pageref{lastpage}}
\maketitle

\begin{abstract}
The stellar mass-to-light ratio gradient in SDSS $r-$band $\nabla (M_*/L_r)$ of a galaxy depends on 
its mass assembly history, which is imprinted in its morphology and gradients of age, metallicity, 
and stellar initial mass function (IMF). Taking a MaNGA sample of 2051 galaxies with stellar masses
ranging from $10^9$ to $10^{12}M_\odot$ released in SDSS DR15, 
we focus on face-on galaxies, without merger and bar signatures, and investigate the dependence of the 2D
$\nabla (M_*/L_r)$ on other galaxy properties, including $M_*/L_r$-colour relationships
by assuming a fixed Salpeter IMF as the mass normalization reference. The median gradient is $\nabla M_*/L_r\sim -0.1$ (i.e., the $M_*/L_r$ is larger at the centre) for massive galaxies, becomes flat around $M_*\sim 10^{10} M_{\odot}$ and change sign to $\nabla M_*/L_r\sim 0.1$ at the lowest masses. 
The $M_*/L_r$ inside a half light radius increases with increasing galaxy stellar mass; in each mass bin, 
early-type galaxies have the highest value, while pure-disk late-type galaxies have the smallest.
Correlation analyses suggest that the mass-weighted stellar 
age is the dominant parameter influencing the $M_*/L_r$ profile, since a luminosity-weighted age is 
easily affected by star formation when the specific star formation rate (sSFR) inside the half light radius 
is higher than $10^{-3} {\rm Gyr}^{-1}$. With increased sSFR gradient, one can obtain a steeper negative $\nabla (M_*/L_r)$. 
The scatter in the slopes of $M_*/L$-colour relations increases with increasing sSFR, for example,
the slope for post-starburst galaxies can be flattened to $0.45$ from the global value $0.87$ in 
the $M_*/L$ vs. $g-r$ diagram. Hence converting galaxy colours to $M_*/L$ should be done
carefully, especially for those galaxies with young luminosity-weighted stellar ages, which can have quite different star formation histories.
\end{abstract}

\begin{keywords}
galaxies: evolution -- galaxies: fundamental parameters -- galaxies: formation
-- galaxies: elliptical and lenticular, cD -- galaxies: spiral -- galaxies: star formation
\end{keywords}



\section{Introduction}
\label{sec:introduction}

The stellar mass assembly history of a galaxy is one of the key parameters for understanding  
its formation and evolution processes.  An important first step is to understand what the 
stellar mass of a galaxy is from the observations we take.  At optical wavelengths, we define
a simple multiplicative relationship between the light received from a galaxy and its mass, the
stellar mass-to-light ratio $M_*/L$.  
Currently, we have three different ways of estimating the mass-to-light ratios 
and thus galaxy stellar masses.

The first method involves performing a stellar population analysis on the observed galaxy spectra or 
broad band spectral energy distributions (SEDs), and calculating the stellar mass based on 
fitted weights to a series of stellar population templates with different stellar 
mass-to-light ratios ($M_*/L$) \citep[see the review by][]{Conroy2013}.

The second converts the galaxy luminosity ($L$) at a specific wavelength band to stellar mass by employing 
an empirical stellar mass-to-light ratio to colour relationship \citep[e.g.][]{Bell2003, 
GB2009, Du2019}.

The third method uses dynamical modelling of a galaxy to obtain $M_*/L$.  For simplicity, the ratio is often assumed 
to be constant over the whole galaxy, and is taken as a free parameter when seeking to reproduce a galaxy's 2D stellar 
kinematic maps \citep[e.g.][]{Cappellari2006, Cappellari2013, Thomas2011, Cappellari2016, Li2017, Lu2020}.

When applying stellar population analysis to obtain a $M_*/L$, the accuracy 
depends on both the fitting algorithm and stellar population models. The empirical $M_*/L$-colour
relation also depends on how well $M_*/L$ can be fitted.
In the galaxy dynamical modelling, the stellar masses are not affected by any uncertainties in 
stellar population analysis, but methods are affected by a stellar and dark matter mass degeneracy 
and the assumption of a constant $M_*/L$ may not be robust.

In the literature, radial $M_*/L$ gradients of galaxies are mainly obtained from stellar population 
analyses of spatially resolved spectra or broad band SEDs by assuming a constant 
initial mass function (IMF). 
For example, \cite{Tortora2011} performed SED fitting to SDSS $ugriz$ bands and found that 
the $M_*/L$ gradients vary with galaxy stellar mass.
Assuming a fixed IMF, the MaNGA work by \citet[]{Li2018} (see their Figure 6) found that the 
$M_*/L$ gradients tend to follow the age gradients: the $M_*/L$ gradient is nearly flat or implies 
a larger $M_*/L$ in the centre for older galaxies and a larger M/L in the outer parts for 
younger ones. Massive elliptical galaxies can also have negative $M_*/L$ gradients,
e.g., \cite{Szomoru2013}, \cite{NET2015}.
\cite{Sonnenfeld2018} obtained gradients by using three different methods: 
$u-g$ colour versus $M_*/L$ relation, $U-B-V$ colours versus $M_*/L$ relation, weak and strong lensing
modelling resulting in three different negative values of $-0.13$, $-0.15$, and $-0.24$, respectively.
CALIFA \citep{Sanchez2012} 
galaxies with morphologies ranging from E0 to Sd types have their $M_*/L$ gradients steeper than 
$-0.2$ in the inner regions and nearly flat in the outer regions \citep{GB2019}.

Galaxy $M_*/L$ gradients can not only be estimated by spatially resolved photometric 
or spectroscopic data, but also predicted using galaxy formation and evolution models.
From theoretical \citep[e.g.][]{WR1978, Hopkins2009, Hopkins2010, Oser2010, Oser2012} 
and observational \citep[e.g.][]{Bezanson2009, Naab2013, MN2018} studies, the formation and 
evolution of elliptical and bulge-dominated spiral galaxies usually occur through 
two distinct phases, i.e., first a ``monolithic'' collapse phase to form the
``in situ'' stars \citep[e.g.][]{ELS1962, Larson1974}, and a second merger-driven growth 
phase to accrete ``ex situ'' stars \citep[e.g.][]{Ciotti2007, Oser2012, RG2016}. 
This two-phase scenario suggests that these galaxies possibly have varying radial gradients 
of stellar population parameters, including stellar age, metallicity, and 
IMF, which are exactly the three parameters that determine a $M_*/L$ from a spectrum. 

For early type galaxies (ETGs), including both ellipticals (E) and lenticulars (S0), 
\cite{Kuntschner2010} and \cite{Li2018} consistently show that the age gradients are nearly flat 
for older ETGs \citep[see also][]{Zheng2017, MN2018}, while younger ETGs tend to have younger cores, 
likely associated to residual star formation in the centres. 
\cite{GD2015} found negative age gradients inside a half light radius (HLR), but nearly flat ones beyond 2 HLR.
Positive age gradients can be obtained by changing the sample and data analysis methods 
\citep[e.g.][]{Koleva2011, Tortora2011, Goddard2017}.
For late type galaxies (LTGs), galaxies with higher stellar masses tend to have steeper negative
age gradients, and those with lower masses have their gradients varying from negative, nearly flat, 
to positive gradients \citep[e.g.][]{Tortora2011, Perez2013, GD2015, Zheng2017}. 

Statistically, ETGs and LTGs have negative metallicity gradients in logarithmic radius, with the values 
ranging from $-0.5$ to 0 \citep[e.g.][]{Mehlert2003, Spolaor2009, Kuntschner2010, GD2015, 
Goddard2017, Zheng2017, Li2018, MN2018, Zibetti2019}.

Evidence of IMF variation has been presented by different authors. Initial convincing evidence 
for an IMF heavier than the Milky Way's in massive ETGs was inferred by modelling stellar 
absorption lines by \cite{vDC2010}. This result appeared consistent with similar evidence from 
mass determinations using strong lensing \citep{Auger2010}. Dynamical modelling of the 
Atlas3D sample by \cite{Cappellari2012} indicated a systematic trend in the IMF, going from 
Milky-Way like in the low velocity dispersion and younger ETGs to Salpeter-like or heavier for the 
high dispersion and older ETGs. A systematic trend was subsequently also inferred from stellar 
population analyses by \cite{Ferreras2013, Spiniello2014, Conroy2017, Li2017, Parikh2018, Vaughan2018b}, 
and \cite{Zhou2019}, although some studies found no clear evidence
\citep[e.g.][]{Zieleniewski2017, Alton2018, Vaughan2018a}.
A recent review of the consistency and tension in IMF determination studies is given by \cite{Smith2020}.

In this paper, we will use SDSS-IV/MaNGA \citep{Bundy2015} IFS data to study $M_*/L$ gradients 
driven by age and metallicity gradients under a fixed Salpeter IMF assumption, and investigate 
how they affect stellar mass estimations and $M_*/L$-colour relations. Our spectral fitting code 
and libraries, and data analysis processes are described in Section 2. We analyze $M_*/L$ gradients 
and $M_*/L$-colour relations for MaNGA galaxies based on fixed IMF assumption 
in Section 3. We compare our results with previous works and discuss the effect of radially 
varying IMFs to $M_*/L$ measurements and $M_*/L$-colour relations in Section 4. 
Our conclusions are summarized in Section 5.

\section{Galaxy sample and data analysis}
\subsection{The galaxy sample selection}
The SDSS 15th data release \citep[DR15,][]{Aguado2019} includes 4672 galaxies with MaNGA IFS observations, 
and also morphological classifications \citep{DS2018} and photometric decompositions \citep{Fischer2019} as well.
These value added catalogues (VACs) allow us to understand how galaxies with different morphologies have evolved. 
We select 2051 face-on viewed (inclination angle $i<45^\circ$) MaNGA galaxies in total by excluding merging and barred galaxies, and those 
with minor and major axes ratio $b/a < 0.5$, with the ratios being taken from \cite{Fischer2019}.

Using galaxy morphologies classified based on deep learning \citep{DS2018} and the photometric decompositions \citep{Fischer2019}, we divide the galaxies we have selected into 
three subsamples to aid our analyses: 1) 873 ETGs 
with Sersic index $n>2.5$; 2) 668 LTGs with both bulge and disk components (bulge+disk LTGs); 
and 3) 510 pure disk LTGs without a bulge component and with $n<2.5$ (pure-disk LTGs).

\subsection{\textsc{pPXF} full-spectrum fitting and the SSP library}

For our selected MaNGA galaxies, we apply the full-spectrum fitting code \textsc{pPXF} \citep{CE2004, Cappellari2017} 
to the galaxies' IFS data. Using this software, when the spectral signal-to-noise ratio (S/N) is larger than 30, 
we can obtain stellar population parameters with biases and scatters less than 0.05 dex \citep{Ge2018}.
We use the version 6.7.6 of the Python code\footnote{Available from https://pypi.org/project/ppxf/} as taken in our previous works \citep{Ge2018,Ge2019} for spectral analyses.

With \textsc{pPXF} selected, an SSP library that can model the evolution of MaNGA galaxies is required.  
\cite{Ge2019} evaluated the three ingredients used for generating an SSP library: the IMF, stellar evolution 
isochrones, and the empirical stellar library.  It was found that local galaxy evolution was best 
described by the Vazdekis/MILES model \citep{Vazdekis2010} with BaSTI isochrones \citep{Pietrinferni2004, Cordier2007}.
For the IMF, it is not possible currently to confirm just how the IMF varies with different galaxies. 
As reviewed in Section 4.2 of \cite{Cappellari2016}, the stellar IMF can vary from a Salpeter IMF
\citep{Salpeter1955} in high mass elliptical galaxies \citep[e.g.][]{Cappellari2012} to a Chabrier IMF 
\citep{Chabrier2003} in low mass spiral galaxies \citep[e.g.][]{Li2017}, with the IMF tending to be 
Kroupa-like \citep{Kroupa2001} in the outskirts of elliptical galaxies \citep[e.g.][]{DS2019}. 
Given that derived stellar population parameters like age, metallicity, and SFR are only  
weakly sensitive to a change in the IMF between Chabrier/Kroupa and Salpeter, we adopt the latter as our 
reference. Any possible radial IMF variation, or IMF variation among galaxies, will produce an offset 
which should be added to the $M_*/L$ values we derive, but is essentially independent of the gradient measurement.
For galaxies with star-forming regions, to cover recent star formation, our youngest age in the SSP is dictated by the limit of the library. Therefore, in this work,
we select a subset of 25 logarithmically-spaced with $\sim 0.11$ dex 
sampled ages between 0.06 and 14 Gyr,  and 12 metallicities 
([M/H]=$-2.27$, $-1.79$, $-1.49$, $-1.26$, $-0.96$, $-0.66$, $-0.35$, $-0.25$, 0.06, 0.15, 0.26, 0.4). Following the data analysis process in the next section, the fraction of spectral fittings with the luminosity-weighted age ($t_L$) younger than 100 Myr is less than $0.04\%$, for which the spectral fitting might be affected due to the existence of stars with ages younger than 60 Myr. This small fraction will not affect our statistical analyses on $M_*/L$ gradients and $M_*/L$-colour relations.

With the \textsc{pPXF} code and SSP library, we derive the stellar population parameters and gas related parameters
separately. For stellar population parameters, as done in \cite{Ge2019}, we perform the \textsc{pPXF} fitting by assuming 
a uniform dust reddening curve given by \cite{Calzetti2000} to correct the intrinsic dust extinction, with all emission
lines masked. For emission line fitting, we re-fit the MaNGA spectra by following the emission line fitting 
example given in the \textsc{pPXF} package: correct the dust extinction of gaseous emissions with Calzetti's dust extinction
curve, but correct the extinction of stellar continuum by adopting a 10-th degree of multi-polynomials (MDEGREE=10), 
by setting the flux ratio of [OI], [OIII] and [NII] doublets fixed at theoretical flux ratio of $\sim 3$, 
[OII] and [SII] doublets restricted to ratios in the physical range. Considering that not all Balmer lines 
are detectable especially for ETGs, we allow free flux fitting of Balmer emission lines, but fix their
line widths to be the same.

\begin{figure*}
\centering
\includegraphics[angle=0.0,scale=0.25,origin=lb]{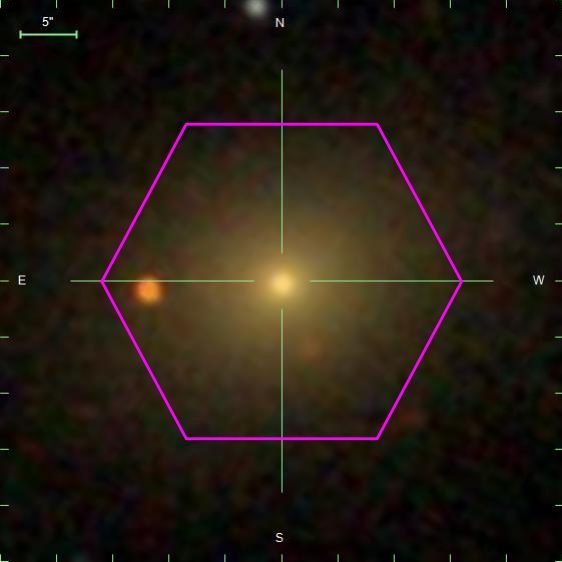}
\includegraphics[angle=0.0,scale=0.25,origin=lb]{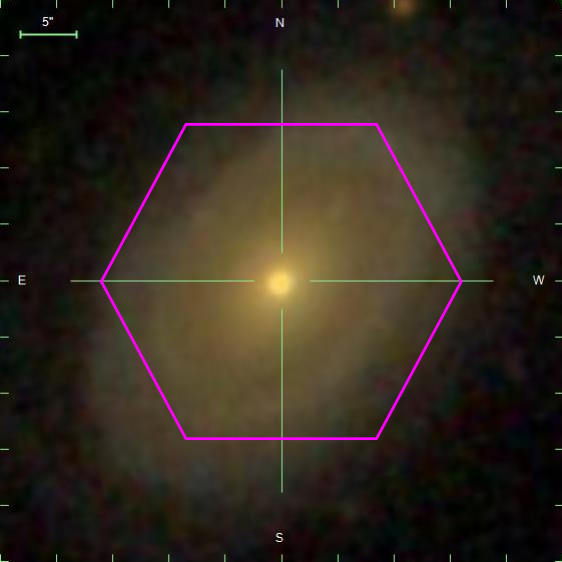}
\includegraphics[angle=0.0,scale=0.25,origin=lb]{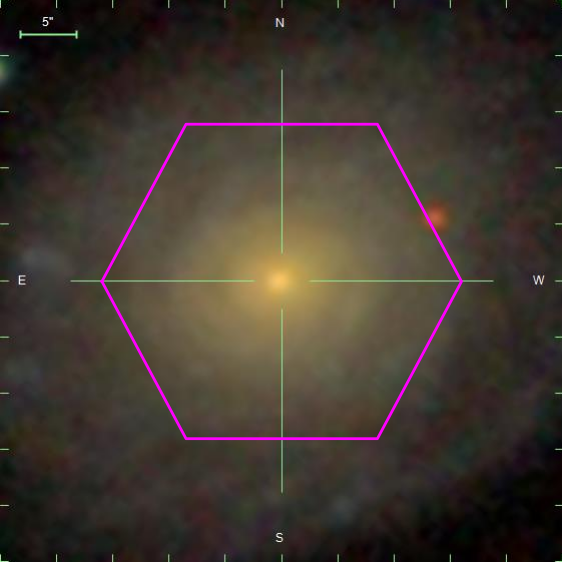}\\
\includegraphics[angle=0.0,scale=0.63,origin=lb]{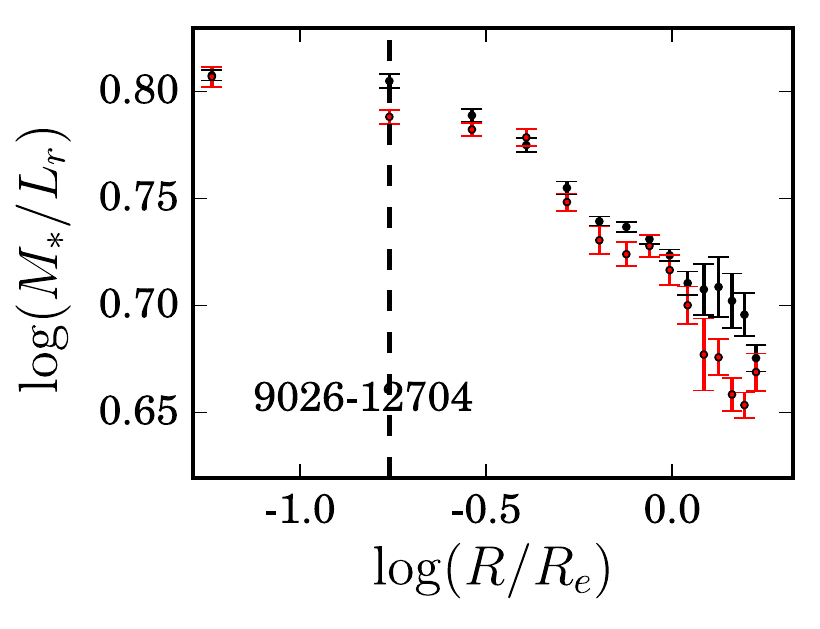}
\includegraphics[angle=0.0,scale=0.63,origin=lb]{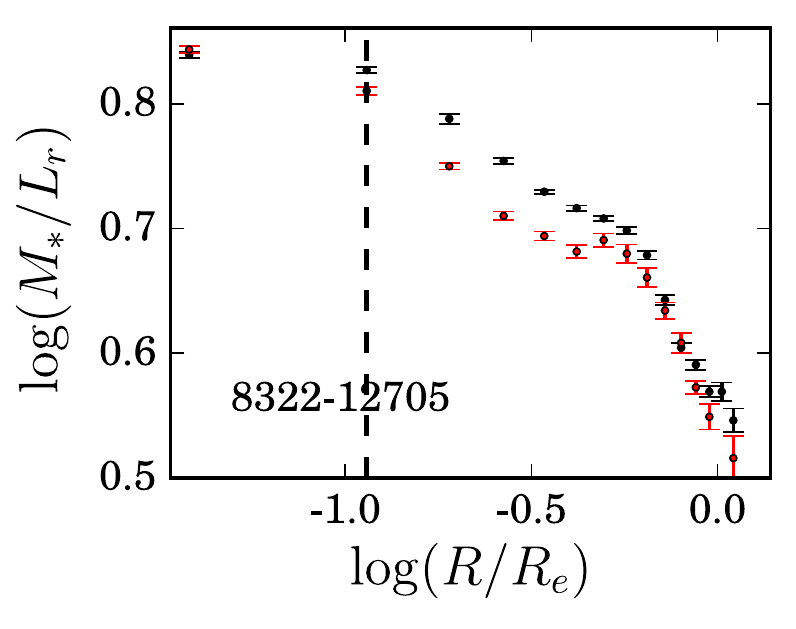}
\includegraphics[angle=0.0,scale=0.63,origin=lb]{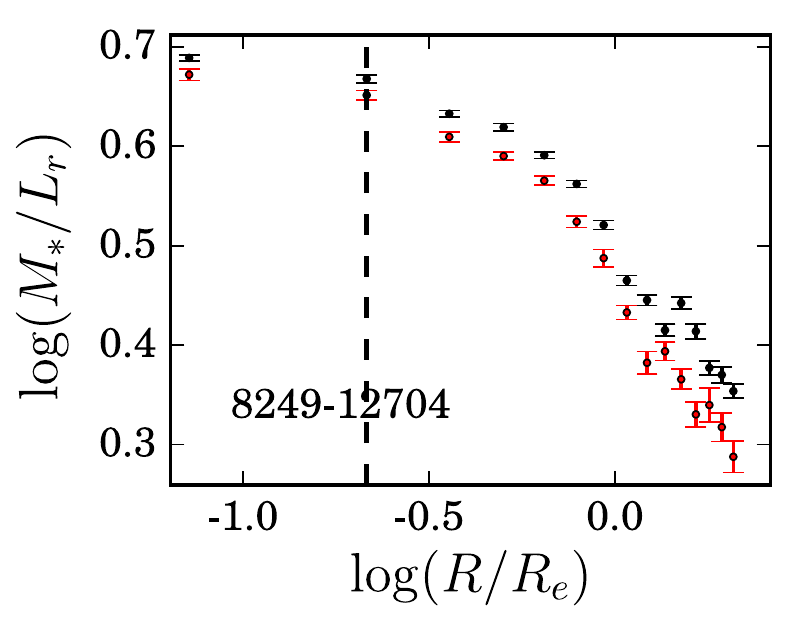}
\caption{Fitting examples of an ETG (left column), a bulge+disk LTG (middle column), 
and a pure-disk LTG (right column). The top row shows the SDSS $gri$-bands stacked image
for each galaxy with the MaNGA FoV overplotted as a pink hexagon. The bottom 
three panels show the corresponding radial $M_*/L_r$ variations of each galaxy, in which 
the black points with error bars are $(M_*/L_r)_{\rm 2D bin}$ with the values calculated from 2D maps, while those
in red represent $(M_*/L_r)_{\rm Rstack}$ with the values obtained from radially stacked spectra. In each bottom panel, the vertical dashed line shows the position of $1.5''$, which is the typical seeing of MaNGA observations.
}
\label{example}
\end{figure*}
\begin{figure*}
\centering
\includegraphics[angle=0.0,scale=1.0,origin=lb]{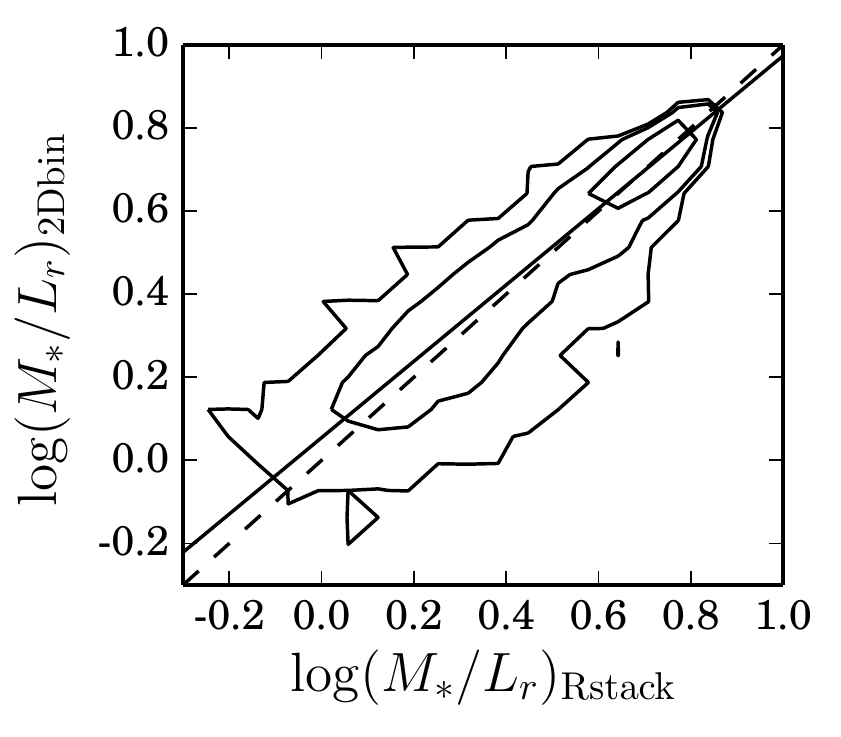}
\includegraphics[angle=0.0,scale=1.0,origin=lb]{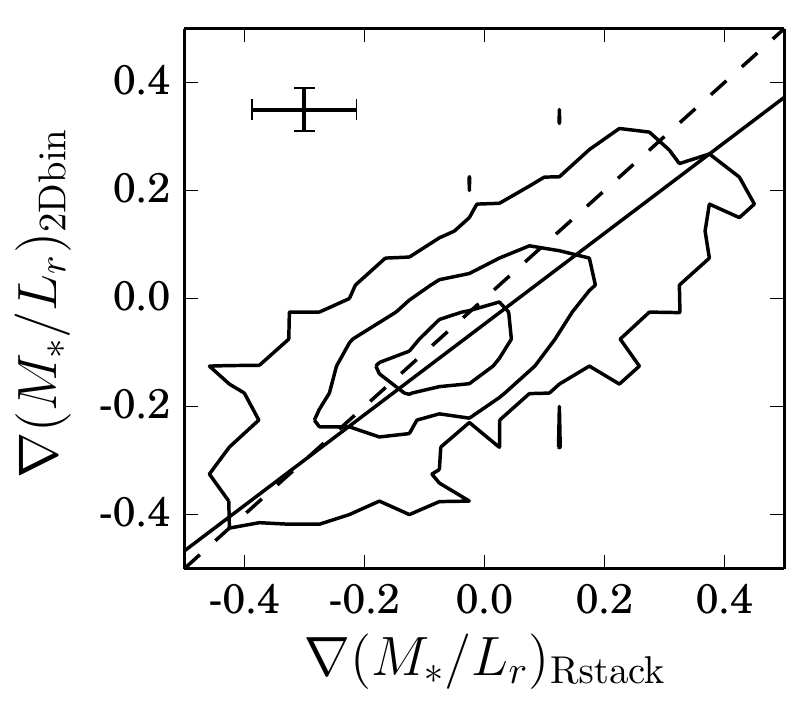}
\caption{Comparison between two representations of $M_*/L_r$ and their radial gradients, with black contours showing the density distributions of Voronoi rebinned spectra in the left panel and galaxies in the right panel, respectively. Left panel: Corresponding to Figure \ref{example}, $\log (M_*/L_r)_{\rm 2D bin}$ are systematically higher than $\log (M_*/L_r)_{\rm Rstack}$ at the low $M_*/L_r$ end, with the linear fitting result $\log (M_*/L_r)_{\rm 2D bin}=0.919\times \log (M_*/L_r)_{\rm Rstack} -0.055$ as shown by the black solid line. Here we use the Python code LTS$\_$LINEFIT \citep{Cappellari2013} 
for correlation analysis, the typical errors of $\log (M_*/L_r)_{\rm 2D bin}$ and $\log (M_*/L_r)_{\rm Rstack}$ are roughly $\sim 0.01$. This correlation has Spearman correlation coefficient $r_s=0.91$
and Pearson correlation coefficient $r_p=0.9$. The linear slope $0.919\pm 0.004$ 
with a scatter of $0.061\pm 0.001$ is
flatter than the dashed diagonal equality line.
Right panel: The $M_*/L_r$ difference between the two methods causes a systematic bias to the measured $M_*/L_r$ gradients, 
with the linear fitting result 
$\nabla (M_*/L_r)_{\rm 2D bin}=0.840\times \nabla (M_*/L_r)_{\rm Rstack} -0.047$
shown by the black solid line. The typical
errors of $\nabla (M_*/L_r)_{\rm Rstack}$ and $\nabla (M_*/L_r)_{\rm 2D bin}$ are displayed
in the top-left of the right panel. This correlation has Spearman correlation coefficient $r_s=0.7$
and Pearson correlation coefficient $r_p=0.73$. The linear slope $0.840\pm 0.013$ 
with a scatter of $0.072\pm 0.004$ is
flatter than the dashed diagonal equality line.
}
\label{mlr_slope}
\end{figure*}
\subsection{IFS data analysis}
Since the surface brightness of a galaxy decreases with increasing radius, current MaNGA 
IFS observations can only provide an SDSS $r$-band, S/N$\sim 1$ per spectral pixel at the edge of 
their field of view (FoV). To improve the robustness of our stellar population analysis 
results, we first select those spaxels with S/N$>2$ and spatially rebin them to S/N$\sim 20$ 
using the Voronoi 2d binning method\footnote{Available from https://pypi.org/project/vorbin/} 
described in \citet{CC2003}. 
There are 690,944 Voronoi bins in total obtained from the 2051 galaxies, with a median of 245 bins for each galaxy, and a median redshift of $z \sim 0.04$, which corresponds to 0.796 kpc/arcsec. For the total 690,944 spectra, $\sim 74\%$ of them have single pixels with $\rm S/N>20$, and only $\sim 3\%$ of them are rebinned from larger than 20 spaxels, which means the diameters of these bins are comparable to or larger than the spatial resolution of MaNGA observations \citep[FWHM=2.5'',][]{Bundy2015}.
When applying the Voronoi 2D binning method to improve the S/N of spaxels with $\rm S/N<20$, the basic assumption for those rebinned spaxels is that they have the same physical properties, since most of them ($\sim 97\%$) have the size smaller than the spatial resolution of MaNGA observation. For each Voronoi bin, we take the mean spectrum of all the stacked spaxels as the stacked spectrum, with the physical parameters of each spaxel in a spatial bin equal to each other.

After applying the \textsc{pPXF} code with our SSP libraries to these spatially rebinned spectra, 
by adopting 10th order multiplicative polynomials, we correct for inaccuracies in the spectral calibration 
and make the resulting data insensitive to reddening by dust \citep{Cappellari2017}.
We then determine the stellar kinematic 2D distributions, which are used subsequently
to correct galaxy rotation during the radial spectral stacking process. 

To study galaxy's $M_*/L$ gradient, we take the $M_*/L$ in the SDSS $r$-band for analysis, with
the definition the same as in Equation (2) of \cite{Ge2018}
\begin{equation}\label{EQ:M/L}
M_*/L_r=\frac{\Sigma f_{M,i}}{\Sigma f_{M,i}/(M_*/L_r)_i}.
\end{equation}
where $M_*$ of the $i$-th SSP template includes the mass in living stars and stellar remnants, 
but excludes the gas lost during stellar evolution. $(M_*/L_r)_i$ corresponds to the $r$-band 
$M_*/L$ of the $i$-th template, and $f_{M,i}$ is the fitted mass fraction.
The IFS spaxels of a galaxy are divided into different 
radial bins based on its ellipticity 
(or the $b/a$ axial ratio), position angle, and the brightest central spaxel in the SDSS $r-$band. 
Considering that the maximum MaNGA FoV of $\sim 30$ arcsec can cover the central 1.5$R_e$ 
for 60 per cent of galaxies and 2.5$R_e$ for 30 per cent of galaxies \citep{Yan2016},
we use the Python package \textsc{MgeFit}\footnote{Available from https://pypi.org/project/mgefit/} by \cite{Cappellari2002}
to model a galaxy's surface brightness within its MaNGA FoV ($\le 30$ arcsec). 
The MGE fitted $b/a$ axial ratio and position angle (rather than values for the 
whole galaxy) are used to construct radial bins for further spectral stacking or parameter 
estimations. 

With interacting and barred galaxies excluded from our sample, the central brightest spaxel 
of each galaxy matches the luminosity-weighted galaxy centre well, and is therefore defined 
as the centre for bin construction. These bins are radial annuli formed by dividing the 
galaxy's major axis into 1 arcsec intervals. Since the pixel size of MaNGA data is 
$0.5\times 0.5$ arcsec, each radial bin includes two spaxels in the major axis and 
at least one spaxel (for $b/a=0.5$) along the minor axis. For each galaxy, there are at most 
15 concentric annuli for studying radial gradients in $M_*/L_r$ and the other stellar 
population parameters. Taking into account the typical seeing of FWHM $\sim 1.5$ arcsec for the 
MaNGA survey \citep[][]{Bundy2015}, we only use those radial annuli whose radii
measured along the major axis are larger than $1.5$ arcsec, and the number of annuli with observed spectra is at least 3 for gradient calculations. Considering that the MaNGA survey is designed for mapping nearby galaxies primarily to $1.5 R_e$ \citep{Yan2016}, we set the cutoff of maximum radii to $1.5 R_e$ for gradient fitting of each galaxy. The minimum radii of the elliptical bins are set to $0.1R_e$ or $1.5''$ for galaxies with low spatial resolution, by taking into account the typical seeing ($1.5''$) of MaNGA observations.

There are two ways of estimating parameters using the radial annuli: either use the median 
parameter values from individual spectra, or stack the spectra and then calculate the parameter 
values. 
For a particular elliptical bin, we can average each parameter based on all the spaxels included in the radial bin. For the second method, we also obtain the mean spectrum of all spaxels in this bin. Therefore, all quantities derived from the two methods can be comparable to each other.
We explain and evaluate both methods, and compare the results obtained.  In the first 
method, after the 2D Voronoi binning, we can perform \textsc{pPXF} fitting on the spectrum associated 
with each Voronoi bin, and calculate the stellar population parameter values, which can then 
be binned into the radial annuli. The radial distribution of each parameter 
(e.g. $(M_*/L_r)_{\rm 2D bin}$) can be obtained by calculating the median values in each annulus, 
with the scatter of each parameter being estimated as the root mean square value.  
In the second method, using the 2D velocity maps determined earlier (with rebinned spectral 
S/N$\sim 20$), we bring all the spectra
in each radial annulus to the same velocity, i.e., $V_*=0$ km/s, and then stack them together.  From these stacked spectra 
in each radial annulus, we can obtain (using \textsc{pPXF} again) the corresponding stellar population 
parameter values (e.g. $(M_*/L_r)_{\rm Rstack}$). 

Figure \ref{example} shows three examples using both methods: an ETG, a bulge+disk LTG, and 
a pure-disk LTG. The bottom three panels show radial variations of $(M_*/L_r)_{\rm 2D bin}$ 
(black colour) and $(M_*/L_r)_{\rm Rstack}$ (red colour), with the corresponding errors estimated in two
different ways. The error bar of $(M_*/L_r)_{\rm 2D bin}$ is calculated based on all the spaxels in 
each annulus by assuming each spaxel has the same $M_*/L_r$ uncertainty. For $(M_*/L_r)_{\rm Rstack}$, 
we obtain its uncertainty from its Monte-Carlo 
based estimation by assuming the flux error of stacked spectra obeys the standard normal distribution. 
For other surveys with higher spatial resolution, e.g., VLT/MUSE \citep{Bacon2014}, one can obtain more radial bins than the MaNGA case, which means more detailed substructures and possibly larger fluctuation appearing in the radial curve compared to MaNGA observations. As to the gradient fitting, the radial elliptical bins taken in our analysis are $2.5$ to $7.5$ times larger than the spatial resolution (FWHM) for the MaNGA primary galaxy sample \citep[Table 3 of][]{Bundy2015}, which can already support a robust gradient measurement.

As shown in Figure \ref{example}, $(M_*/L_r)_{\rm Rstack}<(M_*/L_r)_{\rm 2D bin}$ happens at different
radii for different types of galaxies. ETGs have increasing differences with larger radii; 
bulge+disk LTGs have large differences for $\log(R/R_e)\sim [-1.0, -0.2]$ and no significant difference
at other radii; and pure-disk LTGs show a systematic decrease in $(M_*/L_r)_{\rm Rstack}$ at all 
radii.

To have a thorough understanding on the difference between measured $(M_*/L_r)_{\rm Rstack}$ and $(M_*/L_r)_{\rm 2D bin}$, we compare them directly in the left panel of Figure \ref{mlr_slope}. By applying the Python code LTS$\_$LINEFIT in version 
5.0.18\footnote{Available from  https://pypi.org/project/ltsfit/} \citep{Cappellari2013} for correlation analysis, the fitted slope of the correlation for the $\log(M_*/L_r)_{\rm 2Dbin}$ vs. $\log(M_*/L_r)_{\rm Rstack}$ plot is 
$0.919\pm 0.004$, which is flatter than the dashed diagonal equality line. At 
the high $M_*/L_r$ end, the values derived from the two methods are the same to 
each other, which indicates that for those old spectra without strong SFR, 
both the two methods can converge to the same results. With decreasing 
$M_*/L_r$, the bias of the $M_*/L_r$ measurements increase.

Bias in the two $M_*/L_r$ measurements also introduces a systematic bias to the slopes of radial $M_*/L_r$ gradients for galaxies in our sample as shown in the right panel of Figure \ref{mlr_slope}.
We find that 
$\nabla (M_*/L_r)_{\rm 2D bin}=0.84\times \nabla (M_*/L_r)_{\rm Rstack} -0.05$, 
which is also derived using the Python code LTS$\_$LINEFIT. 
The fitted slope that is flatter than the diagonal equality line should be mainly caused by the 
spatially inhomogeneous surface densities of star formation rate (SFR) inside a galaxy. 
For spaxels in a radial annulus, if their SFRs have large variation, then those 
spaxels with higher SFRs can contribute a larger luminosity fraction than those with lower SFRs due to
the larger luminosity fraction of young and high-mass stars. This makes the spectral fitting
to the stacked spectrum biased to smaller $M_*/L$ since young stellar populations with higher luminosity 
can obscure signals from older ones, hence the derived $(M_*/L_r)_{\rm Rstack}$ is 
smaller than the corresponding $(M_*/L_r)_{\rm 2D bin}$.
To avoid the possible uncertainties caused by SFR variation, in this work, we take 
$(M_*/L_r)_{\rm 2D bin}$ instead of $(M_*/L_r)_{\rm Rstack}$ for gradient and $M_*/L$-colour relation analyses.

The gradients are measured by performing a fit of the linear relation
\begin{equation}\label{def_mlr}
\log(M_*/L_r) = a + b\times \log(R/R_e).
\end{equation}
within the radial range $0.1R_e$ (or 1.5 arcsec for low spatially resolved galaxies)  and $1.5 R_e$. We define the gradient as the slope of the linear fit $\nabla(M_*/L_r)\equiv b$, and perform the fit using Numpy's \citep{Harris2020} \textsc{polyfit}. The formal errors are calculated from the returned covariance matrix.
The gradients of other parameters including luminosity-weighted age $\log(t_L/\rm yr)$, mass-weighted age 
$\log(t_M/\rm yr)$, luminosity-weighted metallicity [M/H]$_L$, mass-weighted metallicity [M/H]$_M$, dust extinction E(B-V), 
and the star formation rate (SFR) and specific SFR in logarithms, i.e., $\log({\rm SFR})$ and $\log({\rm sSFR})$, 
are defined similarly. The errors in these gradients are estimated using the Python numpy \citep{Oliphant2007} 
polyfit routine.

For each spatially rebinned spaxel, its SFR and sSFR are converted from the \textsc{pPXF} fitted $\rm H\alpha$ 
luminosity using the empirical law given by \cite{Kennicutt1998} under the Salpeter IMF assumption,
\begin{equation}
{\rm SFR}(M_\odot{\rm yr^{-1}}) = 7.9\times 10^{-42} L(\rm H\alpha)~ (ergs~ s^{-1}),
\end{equation} 
and the sSFR is defined as
\begin{equation}
{\rm sSFR}~ ({\rm Gyr^{-1}}) = {\rm SFR}/M_*\times 10^9.
\end{equation}
Given that the stellar age, metallicity, SFR, and sSFR of spaxels in ETGs and LTGs can actually cover 
several order of magnitudes, gradients are calculated logarithmically.

\begin{figure}
\centering
\includegraphics[angle=0.0,scale=1.1,origin=lb]{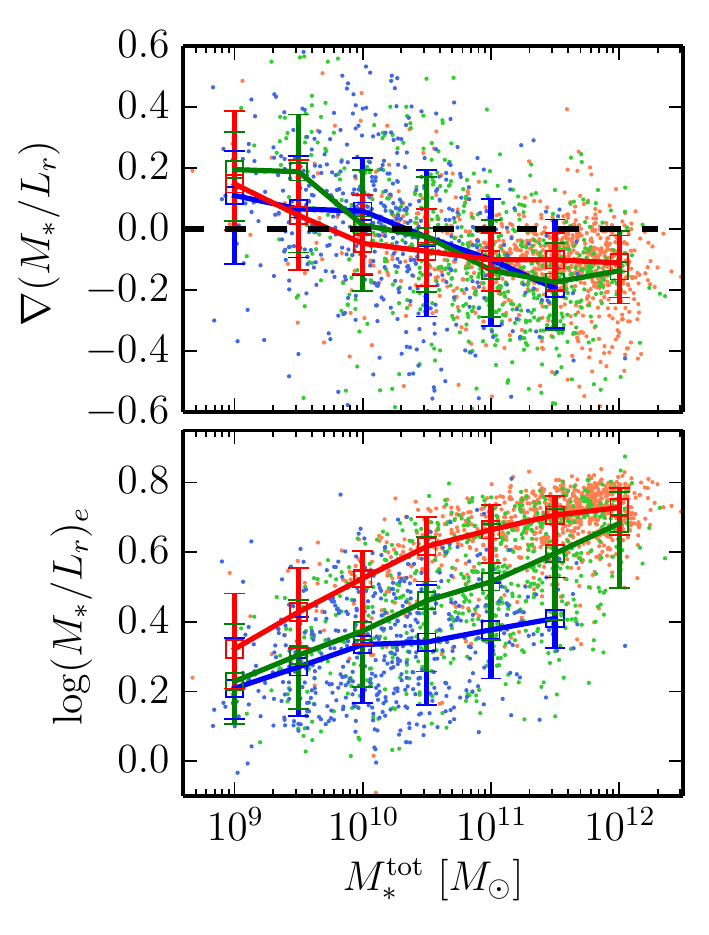}
\caption{The logarithmic gradients of $M_*/L_r$ (top panel) defined in 
Equation (\ref{def_mlr}) and the luminosity weighted mean values of $M_*/L_r$ inside one 
effective radius, i.e., $\log(M_*/L_r)_e$ (bottom panel) as a function of galaxy 
stellar mass with different morphologies. The red, green, and blue points in each panel 
correspond to the ETGs, bulge+disk LTGs, and pure-disk LTGs, respectively. The median 
$M_*/L_r$ gradients in a set of mass bins for the three 
types of galaxies are also shown by open boxes with errors labelled in the same colours 
as the data points. The error bars indicate the 16th and 84th percentiles for each mass bin. The width of each mass bin is 0.5 dex. Only those mass bins with at least 3 galaxies inside are plotted. The dashed line in the top panel shows $\nabla (M_*/L_r)=0$, which helps present when the median gradient reverts from positive to negative.
}
\label{mass_mlr}
\end{figure}
\begin{figure*}
\centering
\includegraphics[angle=0.0,scale=0.9,origin=lb]{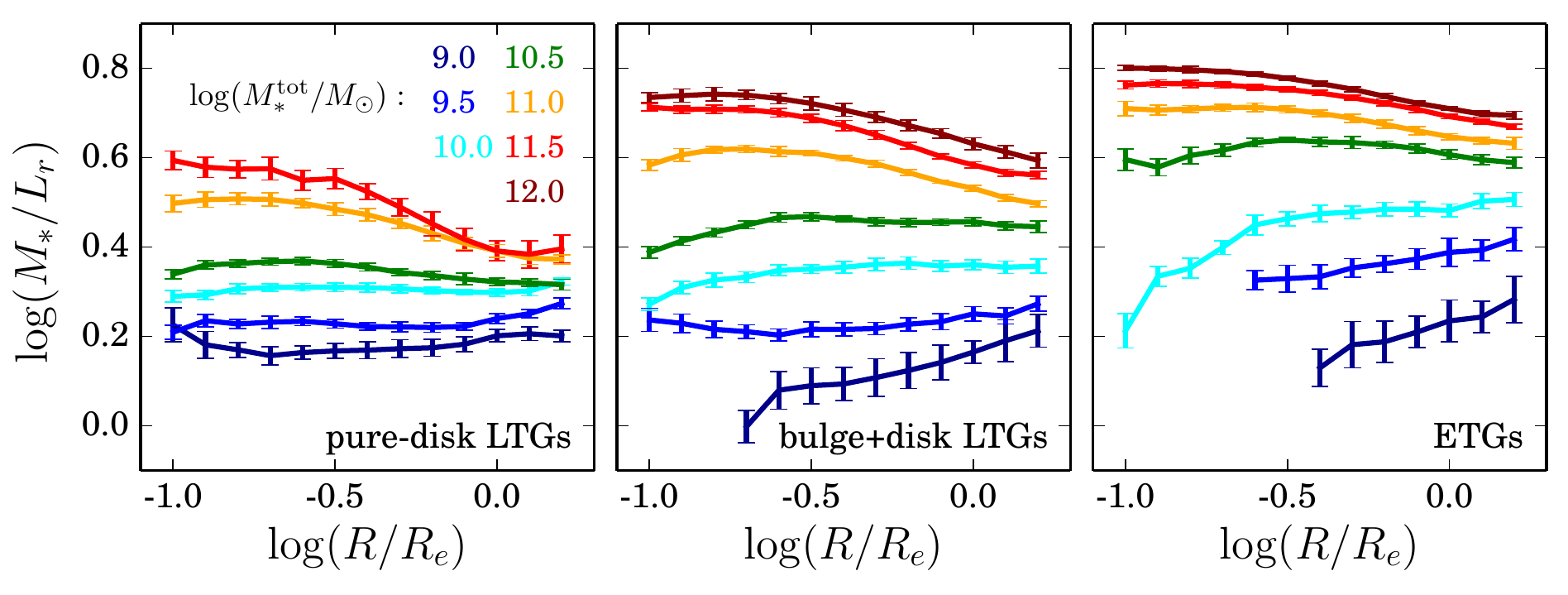}
\caption{The averaged radial $M_*/L_r$ profiles for different galaxy types and mass bins corresponding to Figure \ref{mass_mlr}. Panels from left to right show results for pure-disk LTGs, bulge+disk LTGs, and ETGs. Colours from dark blue to dark red represent the galaxy mass from $\log(M_*^{\rm tot}/M_\odot)=9.0$ to $12.0$, as labelled in the left panel. By setting the galaxy radii ranging from 0.1 to 1.5 $R_e$ evenly sampled logrithmically, with each radial bin having a width of 0.1 dex, we calculate the averaged $\log(M_*/L_r)$ and the corresponding uncertainty only if over 50\% of galaxies included in this bin have been spatially resolved. Considering that the typical seeing of MaNGA observations is $1.5''$, the inner $\sim 3$ data points are mostly affected by the seeing effect than those outside radial bins.
}
\label{global_mlr}
\end{figure*}
\begin{figure*}
\centering
\includegraphics[angle=0.0,scale=0.95,origin=lb]{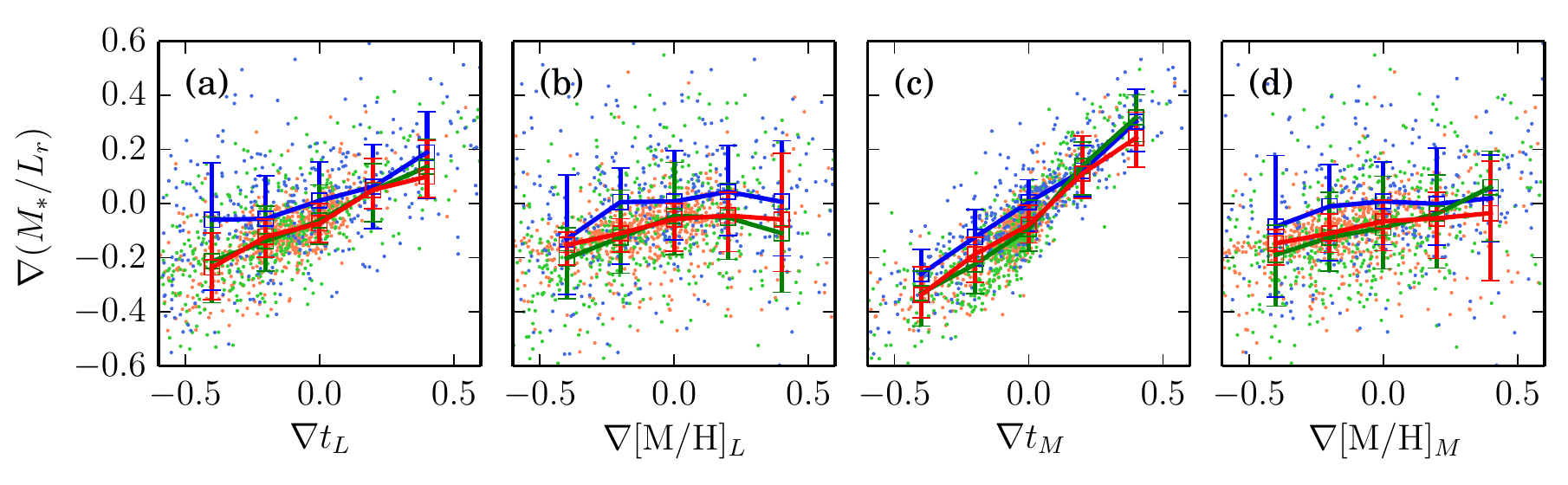}
\caption{Correlations between $M_*/L_r$ gradients and other stellar population parameters: 
gradients of luminosity-weighted stellar age ($t_L$, panel a), luminosity-weighted stellar metallicity 
([M/H]$_L$, panel b), mass-weighted stellar age ($t_M$, panel c), and mass-weighted stellar metallicity
([M/H]$_M$, panel d). ETGs, LTGs with and without a central bulge are shown in red, green, and blue
colours, respectively. The median values in each age or metallicity gradient bins are shown by open boxes 
and the associated error bars (16th and 84th percentiles) in the same colour correspondingly.
}
\label{grad_cor}
\end{figure*}
\begin{figure*}
\centering
\includegraphics[angle=0.0,scale=0.9,origin=lb]{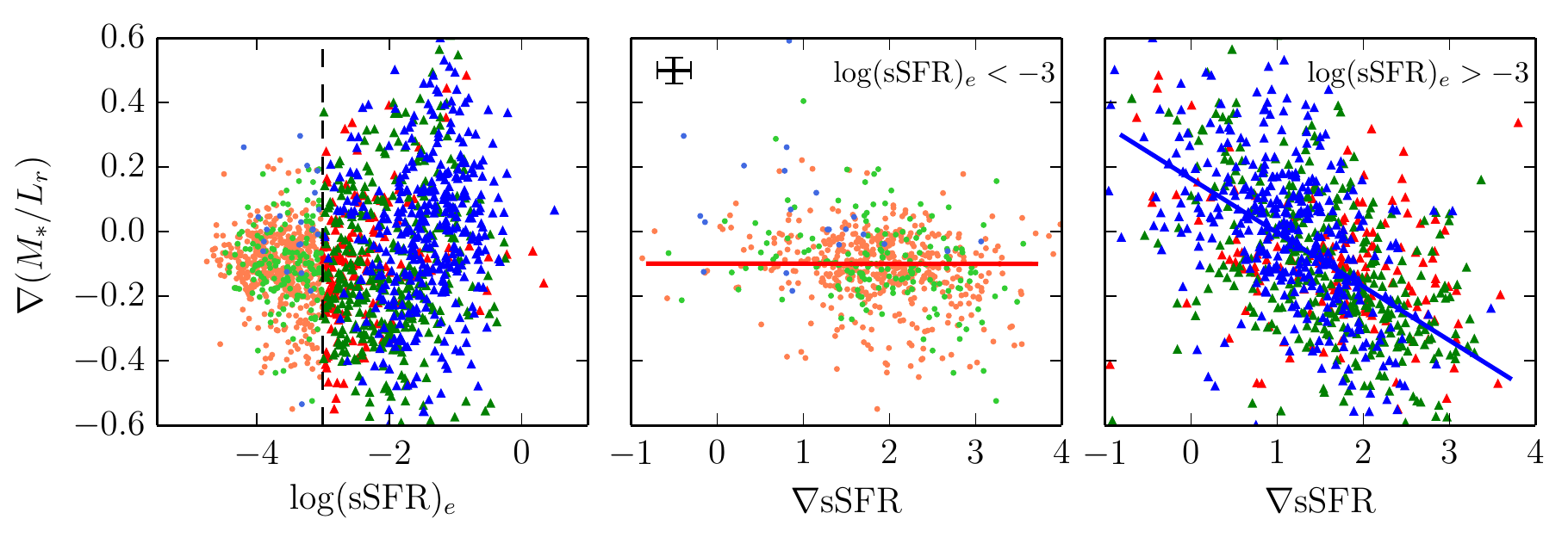}
\caption{The $M_*/L_r$ gradients as a function of sSFR inside a galaxy's half light radius
$\log({\rm sSFR})_e$ (left panel) and $\nabla {\rm sSFR}$ for $\log({\rm sSFR})_e<-3$ (middle panel)
and $\log({\rm sSFR})_e>-3$ (right panel) cases. The red, green, and blue points correspond to the 
ETGs, bulge+disk LTGs, and pure-disk LTGs, respectively. The vertical dashed line in the left panel 
shows the criterion of $\log({\rm sSFR})_e=-3$. Here we especially label those galaxies
with $\log({\rm sSFR})_e>-3$ by triangles in the corresponding darker colours. 
In the middle panel, galaxies with $\log({\rm sSFR})_e<-3$ are plotted 
to present the $\nabla (M_*/L_r)$ as a function of $\nabla {\rm sSFR}$, and a linear fitting
of these points are shown in red solid line with a slope of 0.0. The top-left horizontal and vertical 
error bars show the median errors of $\nabla {\rm sSFR}$ and $\nabla (M_*/L_r)$ of the galaxy sample, respectively.
The right panel plots those galaxies with 
$\log({\rm sSFR})_e>-3$, and has the linear fitting result overlapped as a blue solid line with
a slope of $-0.167\pm0.009$ and scatter of $0.172\pm0.005$ obtained from the LTS$\_$LINEFIT correlation analysis.
}
\label{mlr_ssfr}
\end{figure*}
\begin{figure*}
\centering
\includegraphics[angle=0.0,scale=0.85,origin=lb]{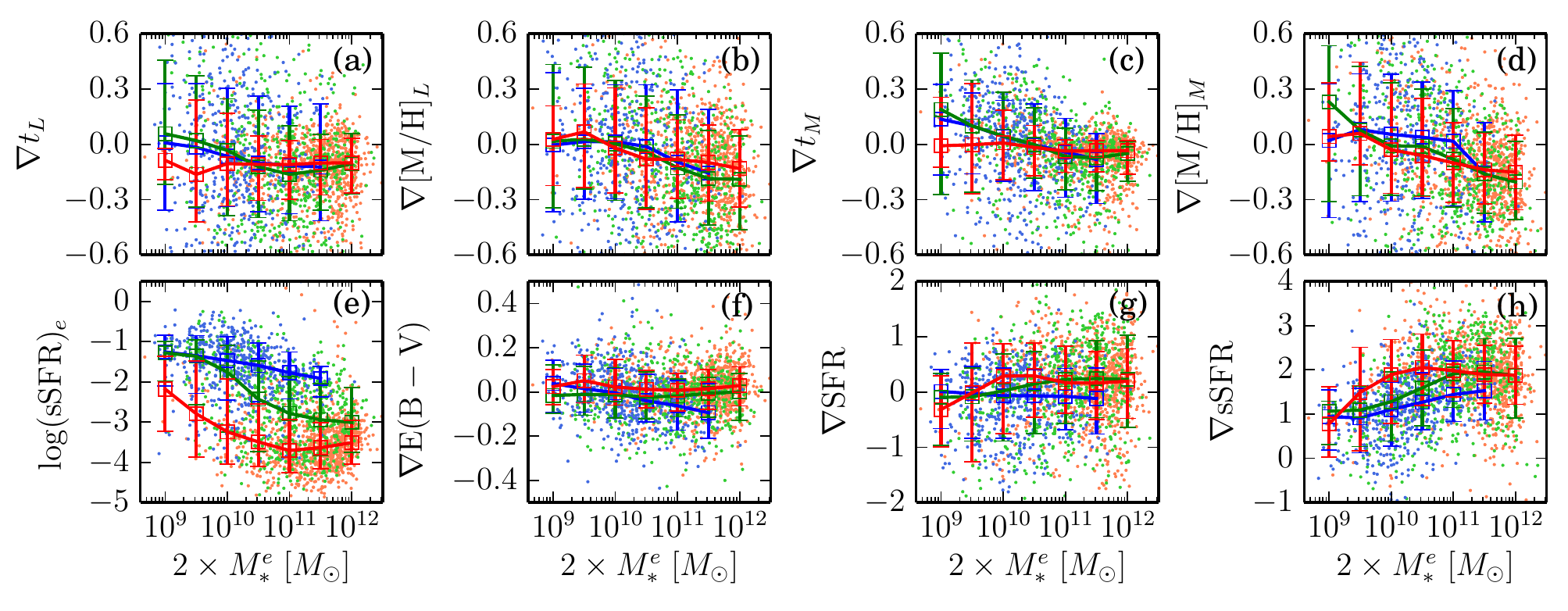}
\caption{Galaxy properties as a function of stellar mass. Panels from (a) to (h)  
show $\nabla t_L$, $\nabla {\rm [M/H]}_L$, $\nabla t_M$, $\nabla {\rm [M/H]}_M$,
$\log({\rm sSFR})_e$, 
$\nabla {\rm E(B-V)}$, $\nabla {\rm SFR}$, and $\nabla {\rm sSFR}$ in terms of the galaxy stellar mass
$M_*^{\rm tot}=2\times M_*^e$. As shown in Figure \ref{mass_mlr}, the red, green, and blue points, and also the 
corresponding median value with error bars represented by 16th and 84th percentiles
show the results of ETGs, LTGs with and without bulges, respectively.
}
\label{grad_n}
\end{figure*}

\section{Results}

We first study the properties of $M_*/L_r$ gradients using a radially fixed Salpeter IMF assumption 
for galaxies with different morphologies.  The results could help understand the evolution of different 
galaxy types and their resulting gradient distributions. We then introduce 
radially varying IMFs and consider their impact on the gradients and the relations between $M_*/L_r$ and galaxy colours.

\subsection{$M_*/L_r$ gradients for galaxies with different morphologies and the fixed Salpeter IMF assumption}

The top panel of Figure \ref{mass_mlr} shows the $M_*/L_r$ gradients as a function of stellar mass 
for three kinds of galaxies: ETGs (red points), bulge+disk LTGs (green points), and pure-disk LTGs 
(blue points). 
Here the galaxy stellar mass is estimated by
\begin{equation}
M_*^{\rm tot} = (M_*/L_r)_e \times L_r^{\rm tot},
\end{equation}
where $(M_*/L_r)_e$ is the projected stellar mass to light ratio inside the half light radius $R_e$ under the assumption of a constant Salpeter IMF, which is calculated based on the 2D $M_*/L_r$ maps and the corresponding $r$-band luminosity
\begin{equation}
(M_*/L_r)_e =  \frac{\sum_{j=1}^N L_{r,j} (M_*/L_r)_j}{\sum_{j=1}^N L_{r,j}}.
\end{equation}
where $N$ is the total number of spaxels inside $R_e$, and $L_{r,j}$ and $(M_*/L_r)_j$ are the $r$-band luminosity and $M_*/L_r$ of the $j$th spaxel, respectively.

This estimate assumes that the $(M_*/L_r)_e$ is representative of the one over the full galaxy, and a better estimate could be obtained if we could directly measure the $M_*/L$ over the full galaxy.
However, the MaNGA survey is limited to 1.5$R_e$ for most galaxies \citep{Bundy2015, Yan2016}, we can not derive the whole 2D mass distribution by only using MaNGA data.
As shown in the top panel,
massive galaxies with total stellar mass larger than $10^{11}M_\odot$ have negative gradients with median values
in the range [-0.2,-0.15], which is consistent with results from \cite{Szomoru2013}, \cite{NET2015}, \cite{Li2018},
and \cite{Sonnenfeld2018}. In particular, \cite{Li2018} used the same approach and MaNGA data but with a different galaxy sample, and our results are consistent with each other. The gradients become shallower with lower stellar mass for galaxies with $M_*^{\rm tot}\le 10^{10}M_\odot$, and this trend is similar to that found in \cite{Tortora2011}. For the $\nabla (M_*/L_r)$-mass 
correlations, all three types of galaxies show no significant difference. Whether pure-disk LTGs 
have a similar formation scenario as elliptical and bulge-dominated galaxies requires further exploration.

The difference between ETGs and bulge+disk LTGs is reflected by the luminosity weighted mean $M_*/L_r$ inside 
the galaxy half light radius $\log(M_*/L_r)_e$ versus galaxy stellar mass plot shown in the bottom panel of Figure \ref{mass_mlr}. 
This correlation is similar to that of $\log(M_*/L_r)_e$ versus velocity dispersion as shown Figure 5 of \cite{Li2018}. With increasing galaxy stellar mass or velocity dispersion, galaxies tend to have higher $M_*/L_r$, in which ETGs have the largest values, 
pure-disk LTGs have the smallest, while those of bulge+disk LTGs lie in between.

Figure \ref{global_mlr} presents another way of analyzing the radial $M_*/L_r$ profiles, by using the averaged radial $M_*/L_r$ profiles for the three kinds of galaxies at different mass bins to study their evolution trends. With increasing stellar mass, the slopes of those radial profiles are positive for low mass bins and become negative for massive bins, presenting similar trends as shown in the top panel of Figure \ref{mass_mlr}. At the same time, the averaged $M_*/L_r$ in each mass bin also increase with increasing galaxy mass, which is consistent with the bottom panel of Figure \ref{mass_mlr}.

To explore the evolution details of these galaxies, we first show $\nabla M_*/L_r$ as a function of 
age and metallicity gradients in both the luminosity- and mass-weighted cases in Figure \ref{grad_cor}.
Since the $M_*/L$ of a spectrum is mainly determined by its stellar age, as a natural consequence, 
$\nabla M_*/L_r$  correlates tightest with the age gradient (panels a and c) rather than the metallicity
gradient as shown in panels (b) and (d). The $M_*/L_r$ gradients correlate with the 
mass-weighted stellar age gradients (panel c) more tightly than the luminosity-weighted ones (panel a).
For all three kinds of galaxies, their $\nabla M_*/L_r$ increases with the stellar mass-weighted age gradient $\nabla t_M$ in a consistent
way and the trends are close to the diagonal line (panel c). For the luminosity-weighted case 
(panel a), the $\nabla M_*/L_r$ vs. the luminosity-weighted age gradient $\nabla(t_L)$ trends are largely biased away from the diagonal line,
with pure-disk LTGs have the larger biases than ETGs and bulge+disk LTGs.
This could be caused by stronger star formation happening in pure-disk LTGs than the
other two kinds of galaxies.

At a certain stellar age, stellar spectra with larger metallicities also have higher $M_*/L_r$ 
\citep[e.g. Figure 2 of][]{Ge2019}. This also explains the positive correlation between $\nabla M_*/L_r$ 
and metallicity gradient $\nabla {\rm [M/H]}$, which is clearly weaker than that between the $\nabla M_*/L_r$ and $\nabla t$, 
as shown in panels (b) and (d) of Figure \ref{grad_cor}.
Compared to the monotonically increasing trend in the mass-weighted case shown in panel (d), 
the weakly increasing trend for $\nabla M_*/L_r$ as a function of $\nabla {\rm [M/H]}_L$ (panel b) 
becomes flat for bulge+disk LTGs and even inverts for pure-disk LTGs when $\nabla {\rm [M/H]}_L > 0$.
Again, this might also be caused by different star formation strengths.

Since star formation can affect the $\nabla (M_*/L_r)$ estimates, 
Figure \ref{mlr_ssfr} is plotted showing the variation in galaxy sSFR with $\nabla (M_*/L_r)$.
Galaxies with $({\rm sSFR})_e < 10^{-3} {\rm Gyr}^{-1}$ 
(solid circles in the left panel) can be classified by the origin of H$\alpha$ emission lines, with ionization not only from SFR, but also hot low-mass evolved stars and weak active galactic nuclei \citep[AGNs, e.g.][]{Stasinska2008, CF2011}. From the WHAN diagram of Cid Fernandes et al. (2011), passive galaxies and LINER dominate the weak emission line systems, which means that galaxies in the middle panel of Figure \ref{mlr_ssfr} are dominated by these two kinds of objects. H$\alpha$ emission lines ionized by hot low-mass evolved stars and weak AGNs would not support a strong correlation between sSFR and $M_*/L_r$ gradients. These passive galaxies with weak SFR have little effect on the $\nabla M_*/L_r$ estimates. Hence we obtain a slope of zero (red line in the middle panel) of their correlations.
For star-forming galaxies with $({\rm sSFR})_e>10^{-3}{\rm Gyr}^{-1}$, a clear anti-correlation between 
$\nabla {\rm sSFR}$ and $\nabla M_*/L_r$ appears (blue line in the right panel). By performing the 
linear correlation analysis with LTS$\_$LINEFIT, we determine that this anti-correlation has a slope of 
$-0.167\pm0.009$ with a scatter of $0.172\pm0.005$.
The two kinds of correlation behaviours indicate that star formation in passive galaxies contributes 
little to the measurements of stellar ages and $M_*/L_r$. However, for star-forming galaxies, a
higher sSFR can produce a younger luminosity-weighted age and hence smaller $M_*/L_r$. This means that 
for a galaxy with a positive sSFR gradient and
$({\rm sSFR})_e>10^{-3}{\rm Gyr}^{-1}$, the galaxy tends to have a more negative $\nabla t_L$ and
$\nabla (M_*/L_r)$. Therefore, the anti-correlation shown in the right panel of Figure \ref{mlr_ssfr}
explains why the $\nabla (M_*/L_r)$ as a function of $\nabla t_L$ presents different trends from that
of $\nabla t_M$ as shown in panels (a) and (c) of Figure \ref{grad_cor}, respectively.

According to the results shown in the bottom panel of Figure \ref{mass_mlr}, bulge+disk LTGs
lie between the ETGs and pure-disk LTGs. In the middle and right panels of Figure \ref{mlr_ssfr}, 
galaxies with different ${\rm sSFR}_e$ have their $\nabla (M_*/L_r)$ and $\nabla {\rm sSFR}$ correlated 
in two ways. To understand further how ETGs and LTGs evolve with their mass, we  
explore the stellar population gradients in terms of their total stellar mass $M_*^{\rm tot}$.

As shown in Figure \ref{grad_n}, both the luminosity-weighted ($\nabla t_L$, panel a) and mass-weighted 
($\nabla t_M$, panel c) age gradients show similar trends with increasing stellar mass as
$\nabla (M_*/L_r)$ (top panel of Figure \ref{mass_mlr}), because of 
their positive correlations as presented in panels (a) and (c) of Figure \ref{grad_cor}. 
The $\nabla {\rm [M/H]}_L$ (panel b) and $\nabla {\rm [M/H]}_M$ (panel d) show a systematic decreasing 
trend with increasing galaxy mass, although with a large scatter when taking the 16th and 84th percentiles 
as error bars. ETGs, with the largest $(M_*/L_r)_e$ (bottom panel of Figure \ref{mass_mlr}), tend to have 
lower sSFR than LTGs, and pure-disk LTGs have the largest sSFR in each mass bin, with bulge+disk LTGs 
lying in between (panel e).
The median values of $\nabla \rm E(B-V)$ (panel f) and $\nabla \rm SFR$ 
(panel g) are consistent with each other and fluctuate around zero. The $\nabla \rm SFR$ distributions
have larger scatters than that of the E(B-V) gradients, especially for ETGs and bulge+disk LTGs. 
Even though we have improved the spectra S/N by spatially stacking spectra, this step mainly improves 
the robustness of E(B-V) estimates resulting from stellar population analysis. The H$\alpha$ 
emission lines are still weak compared to the continua due to the lower SFR of ETGs and bulge-dominant LTGs, 
hence H$\alpha$-based $\nabla \rm SFR$ values show larger scatter than that of $\nabla \rm E(B-V)$.
The age, metallicity, and $M_*/L_r$ gradients have weakly decreasing trends with increasing galaxy mass. 
This can be explained by the correspondingly increased sSFR gradients as shown in panel (h).

\subsection{$M_*/L_r$-colour relations}
For galaxies in our sample, we take the $M_*/L_r$ values for all the spatially rebinned spaxels, and use them to investigate potential relationships between $M_*/L_r$ and colours.
Here we take $g-r$, $r-i$, and $g-i$ colours from the SDSS $gri$ bands for analyses.
By applying the \textsc{pPXF} fitted E(B-V) dust extinction correction to the $gri$ band luminosities, 
all three colours are dust extinction corrected.

\begin{figure*}
\centering
\includegraphics[angle=0.0,scale=0.85,origin=lb]{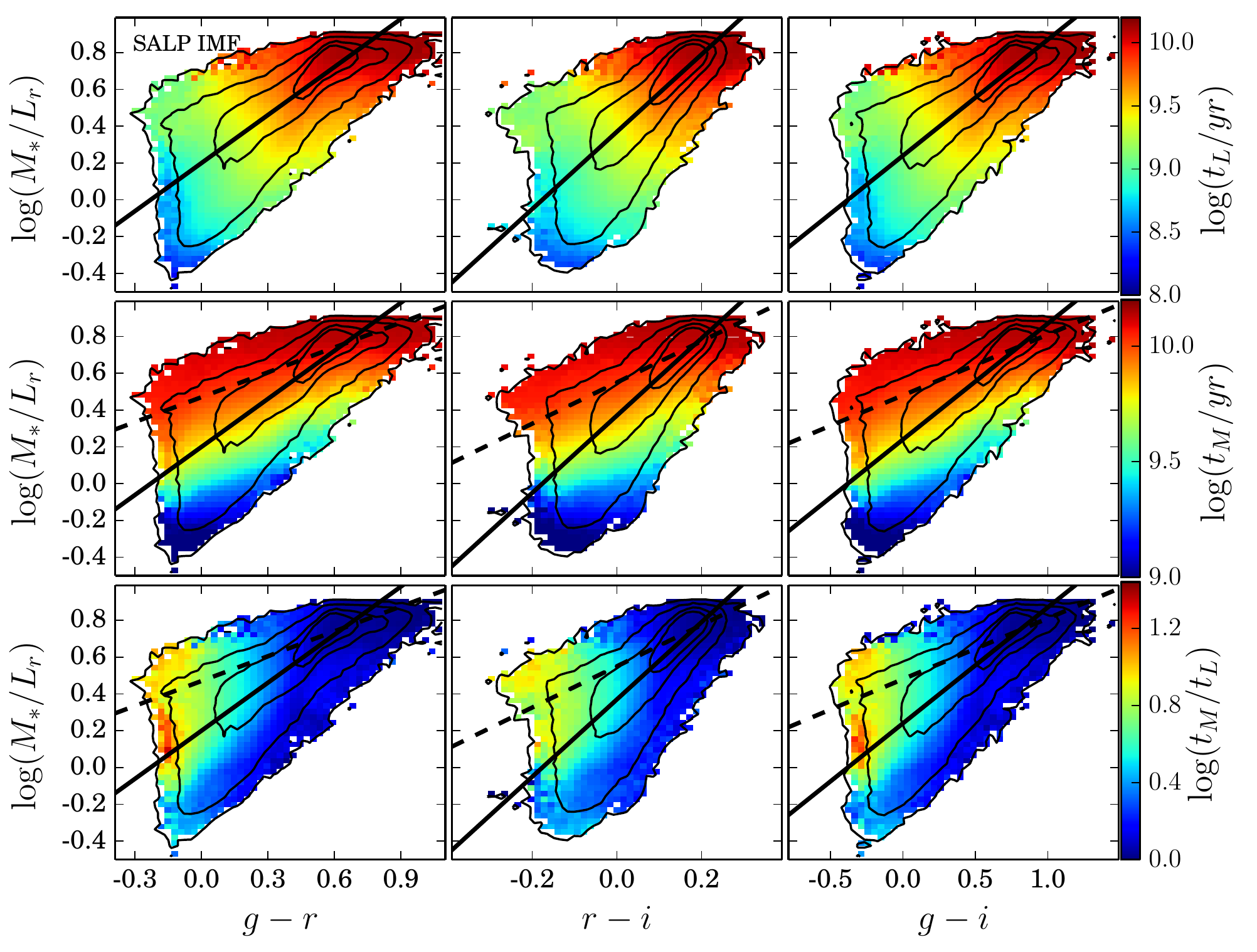}
\caption{The $M_*/L_r$ and colour relations after assuming a universal Salpeter IMF. 
Panels from left to right correspond to the correlations with SDSS $g-r$, $r-i$, and $g-i$ colours, 
with all the colours are dust extinction corrected based on the fitted E(B-V) in \textsc{pPXF} fitting.
In each panel, the black contours show the density distribution of pixels with number 
density larger than 10, which means that all the pixels shown here have S/N$>3$ in the case of 
poisson noise distribution. We fit 
the correlation by $\log(M_*/L_r) = a + b\times {\rm colour}$, and show the 
fitting result as the black solid lines. Those pixels with colours from blue to red represent 
the mean luminosity-weighted (top three panels) or mass-weighted (middle three panels) 
stellar ages from young to old. In the middle three panels, the black dashed lines show
linear fitting results of those spectra with mass-weighted stellar age $t_M>10$ Gyr. In the bottom three panels, we present the age difference between the mass- and luminosity-weighted stellar ages, i.e., $\Delta \log t = \log(t_M/t_L)$. Blue colours indicate SFHs dominated by a single burst while redder colours correspond to more complex SFHs with higher fraction of newly formed stars.}
\label{mlr_colour_salp}
\end{figure*}
\begin{figure}
\centering
\includegraphics[angle=0.0,scale=1.05,origin=lb]{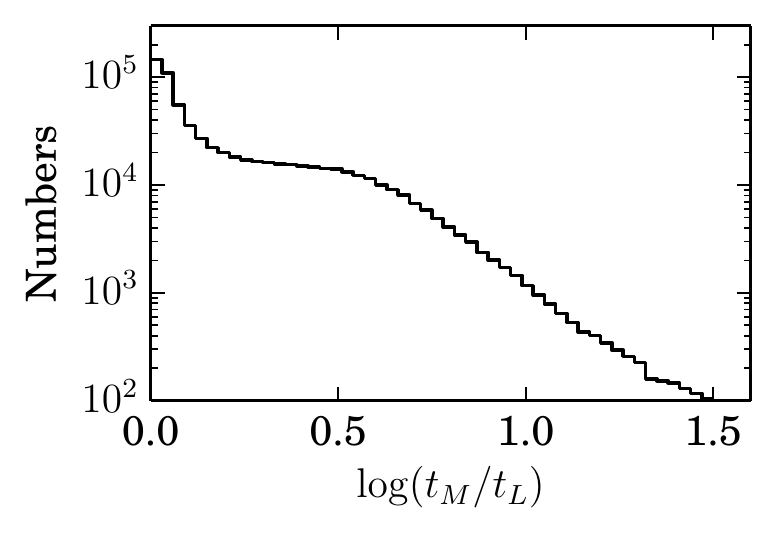}
\caption{Number distribution of $\Delta \log t = \log(t_M/t_L)$ for 690,944 spectra from the Voronoi 2D binning of 2051 galaxies, in which 111,613 spectra ($\sim 16\%$) have $\Delta \log t>0.5$, and 6,777 spectra ($\sim 1\%$) have $\Delta \log t>1.0$.}
\label{dt_psb}
\end{figure}
\begin{table}
\caption{The $M_*/L_r$ as a function of colour at fixed Salpeter IMF assumption}
\begin{center}
\begin{tabular}{|c|c|c|c|c|}
\hline
colour & $a$    & $b$   & $a_{\rm o}$ & $b_{\rm o}$ \\
\hline
$g-r$ & 0.20   & 0.87  & 0.47  & 0.45 \\ 
\hline
$r-i$ & 0.37   & 2.10  & 0.55  & 1.12 \\ 
\hline
$g-i$ & 0.24   & 0.63  & 0.48  & 0.33 \\ 
\hline
\end{tabular}
\end{center}
\small
Note: For the fitting function of $\log(M_*/L_r) = a + b\times {\rm colour}$. 
The subscript $\rm o$ represents the fitting results of those 
spectra with older mass-weighted stellar ages ($t_M>10$ Gyr).
\label{coef_mlr_colour}
\end{table}

Figure \ref{mlr_colour_salp} shows $M_*/L_r$ as a function of SDSS $g-r$ (left column),
$r-i$ (middle column), and $g-i$ colours (right column) under the assumption of a universal Salpeter IMF.
The dashed line in each panel shows the linear fitting result of the $M_*/L_r$-colour correlation,
with the detailed parameters listed in Table \ref{coef_mlr_colour}. In each panel, we can see 
a tight linear correlation \citep[e.g.][]{BelldeJong2001} for $\sim 80$ per cent of spaxels 
(surrounded by the first and second central 
contours). Other works \citep[e.g.][]{GB2019}, find a similar result. 
Compared to \cite{BelldeJong2001}, which studied the $M_*/L$-colour relations by assuming 
different galaxy evolution models, our results provide the statistical $M_*/L$-colour correlation coefficients
of galaxies with real SFHs at redshift range $z\sim [0, 0.15]$ \citep{Bundy2015}. Spectra from IFS observations with 
the FoV covering $\ge 1.5 R_e$ also contain more complex SFHs than single observed spectra only 
focusing on the central 3 arcsec diameter fibre \citep[e.g.][]{Bell2003}. Therefore, the correlation coefficients
of $M_*/L_r$ vs. $g-r, r-i$, and $ g-i$ colour relations vary slightly but are roughly consistent with that of \cite{Bell2003}.

$M_*/L_r$ not only correlates well with galaxy colours, but also with luminosity-weighted stellar ages 
($t_L$, see the top three panels of Figure \ref{mlr_colour_salp}). However, for the mass-weighted 
stellar age ($t_M$) case (the middle row of Figure \ref{mlr_colour_salp}), 
the correlation behaviours among them are different from that of the $t_L$ case.
Spaxels with similar $t_M$ can have 
different colours. This is due to the varied star formation histories that have occurred 
\citep[e.g.][]{Bell2003, GB2009}. As studied in \cite{Yesuf2014} and \cite{Pawlik2018}, post-starburst galaxies can be divided into several types based on the fraction and appearence time of starbursts in the star formation history (SFH). In the bottom row of Figure \ref{mlr_colour_salp}, we plot the age difference of each spectrum $\Delta \log t = \log(t_M/t_L)$ as an indicator for reflecting the contribution of newly formed stars to the whole SFH.
For example, post starburst galaxies with their recent star bursts happening in the last 1 Gyr
can change the galaxy colours from red to blue, but the $t_M$ can still be $\sim 10$ Gyr, 
and is largely biased from an exponential SFH. 
In the bottom panels of Figure \ref{mlr_colour_salp}, those pixels in red show galaxies with SFHs having the strongest recent starbursts.
Spaxels with similar SFHs as post-starburst galaxies can lie on the top-left of the density
distribution in each panel, i.e., spaxels with a little bit smaller $M_*/L_r$, $t_M\sim 10$ Gyr, 
and blue colours, as shown in the middle row of Figure \ref{mlr_colour_salp}.
However, their age differences are larger than those with the same $M_*/L_r$ with redder galaxy colours, or the same galaxy colours with lower $M_*/L_r$.

Those spaxels with large age difference are not insignificant for the current sample. Figure \ref{dt_psb} shows the $\Delta \log t$ distribution for all rebinned spectra of our galaxy sample, in which $\sim 16\%$ of the total spaxels have $\Delta \log t>0.5$ dex, and $\sim 5\%$ of them have $\Delta \log t>1.0$ dex, which means that their newly formed stars can have a significant effect on changing of spectral shape and galaxy colours.

The correlation
coefficients of those spaxels with stellar age older than 10 Gyr are also listed in
Table \ref{coef_mlr_colour}, and are systematically flattened compared to all spaxels. 
A similar result is found by \cite{GB2009} (see their Figure 11), which has less complex
SFHs assumed than the observed MaNGA data. Our results actually complement the missing 
top-left part in their Figure 11 that has blue colours but old populations with high $M_*/L_r$.

With the statistical analyses on the effect of complex SFHs to the $M_*/L_r$-galaxy colour relations, the correlation slopes obtained by different works are mainly determined by their galaxy types. For local galaxies, the slopes obtained in different works are no flatter than the $b_{\rm o}$ listed in Table \ref{coef_mlr_colour}, but the detailed values could vary due to the different selection criteria of their galaxy samples, which indicates that varied SFH distributions are covered in the $M_*/L$-colour relations \citep[e.g.][]{BelldeJong2001, Bell2003, GB2009, GB2019}.

\section{Discussion}

\subsection{Comparison with previous works}

\cite{Tortora2011} performed SED fitting of SDSS $ugriz$ bands for 50,000 galaxies and found that 
galaxies of different types show different behaviours for their $M_*/L$ gradients as a function of 
stellar mass.  For LTGs, gradients steepen negatively with increasing mass, while for ETGs, gradients first
decrease with increasing mass up to $\sim 10^{10.3}M_{\odot}$, and then increase with increasing mass.
The advantage of using photometric data with stellar population analysis is that galaxy images
can be obtained with much less observation time and higher S/N and larger FoV than 
that for IFU observations. However, the fitted results are contaminated by both emission line 
(e.g. H$\alpha$) contributions to different bands (especially for LTGs), and lack of absorption
lines, which introduces large uncertainties caused by the degeneracy between age, metallicity, 
and dust extinction.

Compared to the results obtained from SED fittings \citep{Tortora2011}, MaNGA data provide us with spectral resolution \citep[$R\sim 2000$,][]{Smee2013} high enough to resolve absorption lines
and with wide enough spectral wavelength coverage \citep[3600-10500\AA,][]{Bundy2015} to perform reliable 
stellar population analyses, if we choose a suitable spectral fitting code and SSP library.
For the \textsc{pPXF} code used in this work, the dust extinction uncertainty is less than 0.01 mag for 
galaxy spectra with age $t>0.1$ Gyr, as shown in Figure 4 of \cite{Ge2018}. This indicates that
the degeneracy between stellar age and dust extinction in the SED fitting process cannot contaminate our analyses of MaNGA data if the assumed dust reddening curve is precise enough. However, many efforts focusing on the dust reddening curve show that the dust-star geometry is complex and that star-forming regions even require a two-component dust model for explanation \citep[e.g.][]{Charlot2000, Wilkinson2015, Wilkinson2017, Li2020, Li2021}.

The sample size of \cite{Tortora2011} is, however, larger than the number of galaxies 
selected in our work, and their galaxies are fainter than ours. The photometrically selected galaxy sample can have 
galaxy stellar masses ranging from $10^{7.7}M_{\odot}$ to the most massive galaxies (see their 
Figure 3 for details) in the SDSS photometric survey. The current MaNGA survey has smaller mass
coverage (see Figure \ref{mass_mlr}) than in the Tortora sample.

\subsection{Previous usage of M/L gradients in dynamical models}

In external galaxy dynamical modelling, e.g., Jeans equations modelling \citep{Cappellari2008}, 
orbit based modelling \citep{Schwarz1979}, or particle-based modelling \citep{Syer1996}
in the past it was common to adopt mass-follow-light models assuming a constant $M_*/L$ 
\citep[e.g.][]{vanderMarel1991, Cappellari2006, LM2012}. 
It was motivated by simplicity and by the fact that the dynamical models were targeting 
central parts of early-type galaxies, where population gradients are modest.

Some papers assessed the influence of $M_*/L$ gradients on masses of supermassive black holes 
\citep[e.g.][]{Cappellari2002vmlr, McConnell2013, Thater2017, Thater2019}. 
They found that ignoring $M_*/L$ variations in mass-follow-light models can lead to biases 
in the black hole masses.

However, one fact that is not always appreciated is that the stellar surface brightness 
(as opposed to the stellar mass density) remains the best approximation for the stellar-tracer 
population, even when $M_*/L$ gradients are present. It implies that one can still correctly model 
the galaxies' total density, without distinguishing what fraction is due to the stellar mass and 
which one is due to the dark matter. In this case, one treats the stars only as a tracer population 
orbiting in the total gravitational potential and recovers reliable total densities without the 
need to know the $M_*/L$ gradients \citep[e.g.][]{Cappellari2015, Poci2017, Mitzkus2017}.

On the other hand, focusing on the total density alone, in the presence of significant $M_*/L$ gradients, 
prevents one from measuring unbiased dark matter profiles or estimating the stellar $M_*/L$ (and IMF). 
For this, one has to explicitly include the $M_*/L$ gradients in the mass models as done in more recent 
studies based on IFS data \citep{Mitzkus2017, Poci2017, Li2017}. In this situation, our assessment of 
systematic trends of $M_*/L$ gradients in galaxies becomes relevant.

\begin{figure}
\centering
\includegraphics[angle=0.0,scale=0.85,origin=lb]{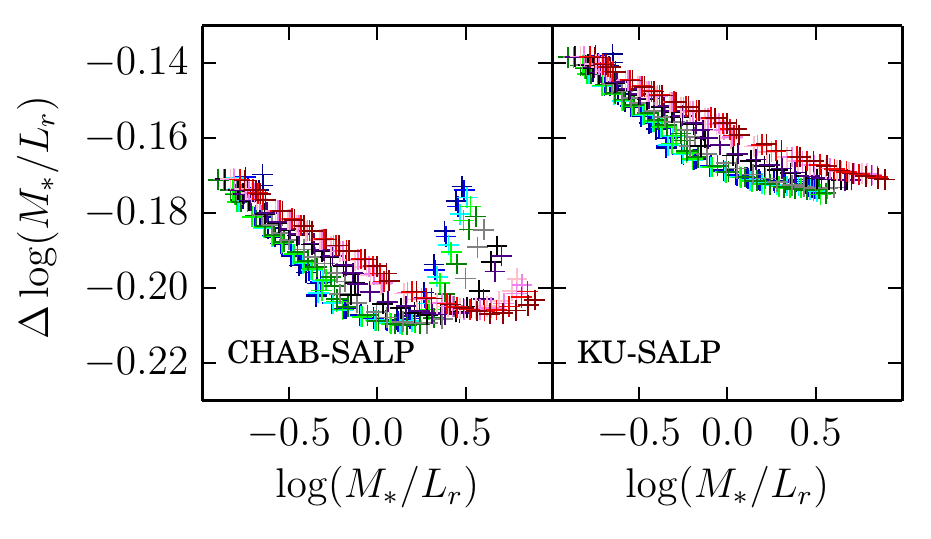}
\caption{The $M_*/L_r$ differences ($\Delta (M_*/L_r)$) as a function of $M_*/L_r$ in the Salpeter IMF for the Vazdekis/MILES model based SSP libraries. The left panel shows the 
$\Delta (M_*/L_r)$ between the Chabrier and Salpeter IMFs (i.e. CHAB$-$SALP), while the 
right panel presents that between Kroupa Universal and Salpeter IMFs (i.e. KU$-$SALP). 
In each panel, the 12 kinds of stellar metallicities from poor ([M/H]$=-2.27$) to rich ($0.4$) 
are labelled with colours from blue to red, respectively.}
\label{ssp_mlr_diff}
\end{figure}

\subsection{Effect of radial IMF variations to $\nabla M_*/L$ measurements}

Once a galaxy has radially varying IMFs \citep[see a review by][]{Smith2020}, then the $M_*/L$ gradients 
will become more negative than the constant IMF case as assumed in this work. Figure \ref{ssp_mlr_diff}
shows the $M_*/L_r$ difference between the Salpeter and Chabrier (or Kroupa) IMFs in the left (or right)
panel, which means that, if we assume a galaxy has a Salpeter IMF at $0.1 R_e$ and a Chabrier (or Kroupa) IMF
at $1 R_e$, the corresponding $M_*/L_r$ difference shown in the left (or right) panel is exactly 
the $M_*/L_r$ gradient that is further steepened \citep[see detailed discussion in][]{GB2019}.

If $M_*/L$-colour relations are applied to convert the galaxy colours to $M_*/L$, the correlation 
slopes are also steepened by potential radial IMF variations, leading to another difference with 
a constant $M_*/L$ approach. For example, \citet{Chao2020} in modelling M87 
utilized a $M_*/L$ profile produced by \citet{Sarzi2018}, which revealed a strong negative IMF 
gradient by taking the IMF-sensitive absorption line features for stellar population analyses,
and caused over a factor 2 increase in $M_*/L$ compared to the case of a Milky Way IMF.

\section{Conclusions}

Using a sample of 2051 face-on galaxies selected from the MaNGA sample released in SDSS DR15, 
we have investigated how galaxy $M_*/L_r$ gradients depend on morphology under the assumption
of a universal Salpeter IMF. We exclude galaxies that are merging, barred, highly inclined ($i>45^\circ$), or have insufficient S/N to ensure robust stellar 
population analyses for calculations of stellar population gradients. We classify our galaxies 
into three groups: 1) ETGs, 2) LTGs with both bulge and disk components (bulge+disk LTGs), 
and 3) LTGs only having one disk component (pure-disk LTGs). 

The $M_*/L_r$ gradients for all the three types of galaxies 
have similar trends as a function of galaxy stellar mass, i.e., $\nabla (M_*/L_r)$ reverts from positive ($\sim 0.1$) to negative ($\sim -0.1$) when galaxy masses increasing from the lowest ($\sim 10^9 M_\odot$) to the highest ones ($\sim 10^{12} M_\odot$). With increasing galaxy stellar mass, the luminosity weighted
$M_*/L_r$ inside a half light radius $\log(M_*/L_r)_e$ also increases, with the trends similar to 
that found in the velocity dispersion $\sigma_e$ vs. $\log(M_*/L_r)_e$ correlations \citep[e.g.][]{Li2018}. 
The age gradients as a function of $M_*^{\rm tot}$ show similar trend as $\nabla M_*/L_r$,
while the metallicity gradients systematically decrease with increasing $M_*^{\rm tot}$.
For $\log(M_*/L_r)_e$ in different mass bin, ETGs have the largest values, and pure-disk LTGs 
have the smallest, while bulge+disk LTGs lie in between.
Correspondingly, the sSFR inside $1 R_e$ (sSFR$_e$) is the lowest for ETGs, and the highest for pure-disk LTGs.

$\nabla (M_*/L_r)$ correlates with mass-weighted stellar age gradients ($\nabla t_M$) more so 
than other parameters. Its correlation with luminosity-weighted age gradients ($\nabla t_L$)
is significantly affected by the star formation when ${\rm sSFR}_e$
is greater than $10^{-3} {\rm Gyr}^{-1}$, where these galaxies have their $\nabla (M_*/L_r)$ 
decreasing with increasing $\nabla {\rm sSFR}$. 
This indicates that a stronger sSFR in the outer radii leads to smaller $t_L$ and $M_*/L_r$, and 
hence more negative $\nabla t_L$ and $\nabla (M_*/L_r)$.
The weak positive correlations between $\nabla (M_*/L_r)$ and metallicity gradients are also 
affected by the galaxy star formation rate.

For the $M_*/L_r$-colour relations, old populations with stellar age older than 10 Gyr 
tend to have shallower correlation slopes than the global ones. In particular, the conversion of $M_*/L$ from galaxy colours
for post starburst galaxies should be very carefully calculated when their SFHs include old populations
dominating the stellar mass but newly formed stars dominating the luminosity.

\section*{Acknowledgements}
We would like to thank to the anonymous referee for the suggestions that helped to improve this paper.
We thank Cheng Li for helpful discussions.
This work is supported by the Beijing Municipal Natural Science Foundation (No. 1204038 to JG), 
by the National Key Research and Development Program of China 
(No. 2018YFA0404501 to SM), by the National Natural Science Foundation of
China (NSFC) under grant numbers 11903046 and U1931110 (JG), 11333003 and 11761131004 (SM), and 11390372 (SM, YL),
and by the National Key Program for Science
and Technology Research and Development (Grant No. 2016YFA0400704 to YL).
RY acknowledges support by National Science Foundation grant AST-1715898.     

Funding for the Sloan Digital Sky 
Survey IV has been provided by the 
Alfred P. Sloan Foundation, the U.S. 
Department of Energy Office of 
Science, and the Participating 
Institutions. 

SDSS-IV acknowledges support and 
resources from the Center for High 
Performance Computing  at the 
University of Utah. The SDSS 
website is www.sdss.org.

SDSS-IV is managed by the 
Astrophysical Research Consortium 
for the Participating Institutions 
of the SDSS Collaboration including 
the Brazilian Participation Group, 
the Carnegie Institution for Science, 
Carnegie Mellon University, Center for 
Astrophysics | Harvard \& 
Smithsonian, the Chilean Participation 
Group, the French Participation Group, 
Instituto de Astrof\'isica de 
Canarias, The Johns Hopkins 
University, Kavli Institute for the 
Physics and Mathematics of the 
Universe (IPMU) / University of 
Tokyo, the Korean Participation Group, 
Lawrence Berkeley National Laboratory, 
Leibniz Institut f\"ur Astrophysik 
Potsdam (AIP),  Max-Planck-Institut 
f\"ur Astronomie (MPIA Heidelberg), 
Max-Planck-Institut f\"ur 
Astrophysik (MPA Garching), 
Max-Planck-Institut f\"ur 
Extraterrestrische Physik (MPE), 
National Astronomical Observatories of 
China, New Mexico State University, 
New York University, University of 
Notre Dame, Observat\'ario 
Nacional / MCTI, The Ohio State 
University, Pennsylvania State 
University, Shanghai 
Astronomical Observatory, United 
Kingdom Participation Group, 
Universidad Nacional Aut\'onoma 
de M\'exico, University of Arizona, 
University of Colorado Boulder, 
University of Oxford, University of 
Portsmouth, University of Utah, 
University of Virginia, University 
of Washington, University of 
Wisconsin, Vanderbilt University, 
and Yale University.

\section*{Data availability}

The data underlying this article will be shared on reasonable request to the corresponding author.











\bsp	
\label{lastpage}
\end{document}